\documentclass[5p,twocolumn,twoside,times]{elsarticle}

\usepackage{ecrc}

\volume{00}

\firstpage{1}

\journalname{Physics Letters A}

\runauth{D.\ S.\ Goldobin}

\usepackage{graphicx}
\usepackage{amsmath}
\usepackage{amssymb}
\usepackage{mathtext}

\begin{document}

\begin{frontmatter}

\title{Anharmonic resonances with recursive delay feedback}

\author{Denis S.\ Goldobin}
\address{Department of Mathematics, University of Leicester,
             Leicester LE1 7RH, UK}
\address{Institute of Continuous Media Mechanics, UB RAS,
             Perm 614013, Russia}
\ead{Denis.Goldobin@gmail.com}
%

\begin{abstract}
We consider application of time-delayed feedback with infinite
recursion for control of anharmonic (nonlinear) oscillators
subject to noise. In contrast to the case of a single delay
feedback, recursive delay feedback exhibits resonances between
feedback and nonlinear harmonics, leading to a resonantly strong
or weak oscillation coherence even for a small anharmonicity.
Remarkably, these small-anharmonicity induced resonances can be
stronger than the harmonic ones. Analytical results are confirmed
numerically for van der Pol and van der Pol--Duffing oscillators.

\vspace{2mm}
\end{abstract}

\begin{keyword}
Noisy oscillators \sep Recursive delay feedback \sep Control \sep
Anharmonicity

\vspace{3mm}

\PACS
 05.40.-a    
\sep
 02.50.Ey    
\end{keyword}

\end{frontmatter}

\section{Introduction}
Delayed feedback is an effective control tool for chaotic and
stochastic dynamics of solely oscillatory
systems~\cite{Pyragas-1992, Just-Benner-Schoell-2003,
Goldobin-Rosenblum-Pikovsky-2003, Boccaletti-Allaria-Meucci-2004,
Janson-Balanov-Schoell-2004, Pomplun-Amann-Schoell-2005,
Goldobin-2008} and collective dynamics of ensembles of
oscillators~\cite{Rosenblum-Pikovsky-2004a,
Rosenblum-Pikovsky-2004b, Popovych-Hauptmann-Tass-2006,
Goldobin-Pikovsky-2006b}. For a recursive delay feedback ({\it or}
``extended delay feedback'') the forcing on the system at the time
moment $t$ is determined not merely by the state difference
 ${\bf z}(t)=[{\bf x}(t-\tau)-{\bf x}(t)]$
($\tau$ is the delay time) but by a chain of differences
 ${\bf z}(t)=\sum_{n=0}^{+\infty}R^n[{\bf x}(t-(n+1)\tau)-{\bf x}(t-n\tau)]$
with $|R|<1$ \cite{Socolar-Sukow-Gauthier-1994,Pyragas-1995}. The
employment of such a feedback can provide new opportunities in
comparison to the case of the ``single delay feedback''
control~\cite{Socolar-Sukow-Gauthier-1994, Pyragas-1995,
Ahlborn-Parlitz-2004, Pawlik-Pikovsky-2006}. Moreover, the
feedback has such a form when the system is connected to a
resonator (for instance, an optical delayed feedback for lasers is
realized with a Fabry–-Perot interferometer). A recursive delay
feedback is also constructively inherent to some electronic
devises (e.g.~\cite{Ryskin_etal-2004}). It is noteworthy that,
from the mathematical perspective, the infinite recursion is
equivalent to the addition of the neutral equation
 ${\bf s}(t)=R{\bf s}(t-\tau)+{\bf x}(t-\tau)$
(then the feedback signal ${\bf z}(t)={\bf s}(t)-{\bf x}(t)$).
Properties of this neutral equation provide favourable conditions
for resonant phenomena in the system.

For limit-cycle oscillators subject to noise, the regularity of
dynamics is characterized by oscillation
coherence~\cite{Pikovsky-Rosenblum-Kurths-2001} and
``reliability''~\cite{Mainen-Sejnowski-1995}. Coherence can be
quantified by the diffusion constant of the oscillation phase
$\varphi(t)$. While for vanishing noise the phase grows linearly
with time, noise results in random excursions of the phase around
the average growth trend,
 $\langle[\varphi(t)-\varphi(0)-\langle\dot\varphi\rangle t]^2\rangle\approx Dt$,
where $D$ is the diffusion constant. The ``reliability'' is the
ability of a system to provide one and the same response for the
noise signal of a prerecorded waveform. Mathematically,
reliability means stability of a system response and can be
measured by the Lyapunov exponent~\cite{Pikovsky-1984, Ritt-2003,
Teramae-Tanaka-2004, Goldobin-Pikovsky-2005a,
Goldobin-Pikovsky-2005b}. The effect of a control on the dynamics
regularity can be, therefore, tracked with the phase diffusion
constant and the Lyapunov exponent.

The employment of single and/or recursive delay feedback control
received previously analytical treatments only for the case of a
harmonic oscillator~\cite{Goldobin-Rosenblum-Pikovsky-2003,
Goldobin-2008, Pawlik-Pikovsky-2006}. In this Letter, the
analytical treatment is generalized for anharmonic oscillators.
Strong anharmonic resonances are revealed even for a relatively
small anharmonicity.

The paper is organized as follows. In Sec.~\ref{sec2} we derive
algebraic equations for the mean frequency, the diffusion
constant, and the Lyapunov exponent for general limit-cycle
oscillator subject to noise and a general linear feedback
represented in terms of a Green's function. In Sec.~\ref{sec3} we
consider the form of these equations specific to a recursive delay
feedback and treat the role of anharmonicity. The analytical
theory we construct is applied to van der Pol and van der
Pol--Duffing oscillators. The anharmonic resonances are also
observed with a direct numerical simulation. Finally, conclusions
are summarized in Sec.~\ref{conclusion}.
\vspace{3mm}

\section{Phase description}\label{sec2}
Let us consider an $N$-dimensional limit-cycle oscillator subject
to recursive delay feedback and noise:

\begin{equation}
\dot{x}_i=F_i({\bf x})+a\,z_i(t)+B_i({\bf x})\circ\xi(t),
 \label{eq-01}
\end{equation}
where $i=1,2,...,N$, $a$ is the feedback strength, $az_i(t)$ is
the feedback term, $\xi(t)$ is white Gaussian noise,
$\langle\xi\rangle=0$ and
$\langle\xi(t)\xi(t')\rangle=2\delta(t-t')$, ``$\circ$'' indicates
the Stratonovich form of stochastic equation. For briefness, we
will first keep the feedback term in a general form valid for any
linear feedback (as in
Refs.~\cite{Tukhlina_etal-2008,Goldobin-2011a})
\[
z_i(t)=\int\limits_0^{+\infty}\sum_{j=1}^{N}G_{ij}(t_1)\,x_j(t-t_1)\,dt_1\,,
\]
involving Green's function $G_{ij}(t)$; for recursive time-delayed
feedback
\begin{eqnarray}
{\bf\hat G}(t)
 =2{\bf\hat S}\sum\limits_{n=0}^{+\infty}\!R^n
 \big[\delta(t-(n+1)\tau)-\delta(t-n\tau-0)\big]\quad
\nonumber\\[5pt]
 =2{\bf\hat S}\Bigg[-\delta(t-0)
 +(1-R)\sum\limits_{n=1}^{+\infty}\!R^{n-1}\delta(t-n\tau)\Bigg]\,,
\label{eq-02}
\end{eqnarray}
where ${\bf\hat S}$ is a constant matrix, $\tau$ is the delay
time, $|R|<1$, and we explicitly indicate that
$\int_0^{+\infty}\delta(t-0)\,dt=1$.

For a noise-free limit-cycle oscillator, the oscillation phase
$\varphi=\varphi({\bf x})$ can be introduced on the limit cycle
${\bf x}_0(t)={\bf x}_0(t+T_0)$ and within its finite vicinity in
such a way that $\dot\varphi=\Omega_0=2\pi/T_0$, where $\Omega_0$
is the natural frequency of the
oscillator~\cite{Winfree-1967,Kuramoto-2003}. The limit cycle can
be parameterized with the phase, ${\bf x}_0(\varphi)={\bf
x}_0(\varphi+2\pi)$. In the presence of a weak forcing (noise and
feedback) the phase description can be still utilized;
\begin{eqnarray}
\dot{\varphi}=\Omega_0+a\int\limits_0^{+\infty}
 \sum_{i=1}^{N}\sum_{j=1}^{N}G_{ij}(t_1)\,H_{ij}\big(\varphi(t-t_1),\varphi(t)\big)\,dt_1
\nonumber\\
 +\varepsilon f\big(\varphi(t)\big)\circ\xi(t)\,,\quad
 \label{eq-03}
\end{eqnarray}
where
\[
f(\varphi_1):=\sum_{j=1}^N
 \left(\frac{\partial\varphi({\bf x})}{\partial x_j}B_j\right)_{{\bf x}={\bf x}_0(\varphi_1)}
\]
is a $2\pi$-periodic function featuring the sensitivity of the
phase to noise, $\varepsilon$ is the noise amplitude,
$H_{ij}(\psi,\varphi)$ is the increase of the phase growth rate
created by the feedback term $x_j(\psi)$ acting on the variable
$x_i(\varphi)$;
\[
H_{ij}(\psi,\varphi_1):=
 \left.\frac{\partial\varphi({\bf x})}{\partial x_i}\right|_{{\bf x}={\bf x}_0(\varphi_1)}
 x_j(\psi)\,.
\]
Notice, accurate derivation of
Eq.~(\ref{eq-03})~\cite{Yoshimura-Arai-2008,
Teramae-Nakao-Ermentrout-2009, Goldobin_etal-2010} yields a drift
term $\propto\varepsilon^2$, which represents the role of the
amplitude degrees of freedom. Since Eq.~(\ref{eq-03}) provides a
noise-induced mean frequency shift $\propto\varepsilon^2$, i.e.,
of the same order of magnitude, this omitted term should be
accounted for, when one considers the effect of noise on the mean
frequency. However, for robustness quantifiers (the phase
diffusion constant and the Lyapunov exponent, which are of our
interest here) this term has been shown to be
negligible~\cite{Goldobin_etal-2010}.

For weak noise, the linear in noise approximation is relevant and
yields (see Appendix for detail) the mean frequency
\begin{equation}
\Omega=\Omega_0+a\int_0^{+\infty}
  \sum_{i=1}^{N}\sum_{j=1}^{N}G_{ij}(t)\,h_{ij}(-\Omega t)\,dt\,,
\label{eq-04}
\end{equation}
the phase diffusion constant
\begin{eqnarray}
 D=\frac{2\varepsilon^2\langle{f^2}\rangle_\varphi}{\Big(1+a\int\limits_0^{+\infty}
 t\sum\limits_{i=1}^{N}\sum\limits_{j=1}^{N}G_{ij}(t)\,h'_{ij}(-\Omega t)\,dt\Big)^2}
\quad
\nonumber\\
 =2\varepsilon^2\langle{f^2}\rangle_\varphi\left(\frac{\partial\Omega}{\partial\Omega_0}\right)^2,
\label{eq-05}
\end{eqnarray}
and the leading Lyapunov exponent
\begin{eqnarray}
\lambda=-\frac{\varepsilon^2\langle(f')^2\rangle_\varphi}{\Big(1+a\int\limits_0^{+\infty}
 t\sum\limits_{i=1}^{N}\sum\limits_{j=1}^{N}G_{ij}(t)\,h'_{ij}(-\Omega t)\,dt\Big)^2}
\quad
\nonumber\\
 =-\varepsilon^2\langle(f')^2\rangle_\varphi\left(\frac{\partial\Omega}{\partial\Omega_0}\right)^2,
\label{eq-06}
\end{eqnarray}
where
$\langle...\rangle_\varphi\equiv(2\pi)^{-1}\int_0^{2\pi}...d\varphi$,
\[
h_{ij}(\psi):=\langle H_{ij}(\varphi+\psi,\varphi)\rangle_\varphi,
\]
is the average susceptibility of the phase to the feedback term
$x_j(\varphi+\psi)$ acting on $x_i(\varphi)$ (notice, this means a
phase delay by $-\psi$), and the prime denotes derivative.
Eq.~(\ref{eq-04}) can be solved with respect to $\Omega$, and
then, with $\Omega$ evaluated, Eqs.~(\ref{eq-05}) and
(\ref{eq-06}) provide $D$ and $\lambda$. Noticeable constancy of
the ratio
\[
\frac{-\lambda}{D}=\frac{\langle(f')^2\rangle}{2\langle{f^2}\rangle}
\]
was already discussed in the literature~\cite{Goldobin-2008,
Goldobin-2011a, Pomplun-Amann-Schoell-2005}. Henceforth, we do not
consider $\lambda$ because considering the diffusion constant is
enough.

\begin{figure}[!t]
\center{\sf
\begin{tabular}{c}
(a)\hspace{-4mm}\includegraphics[width=0.190\textwidth]%
 {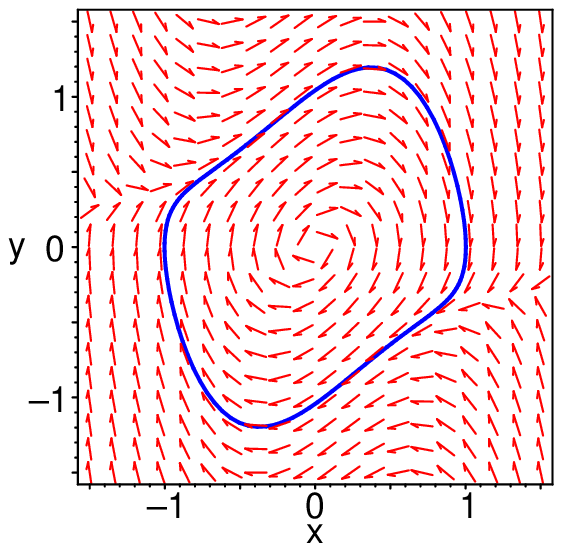}
 \quad
(b)\hspace{-4mm}\includegraphics[width=0.248\textwidth]%
 {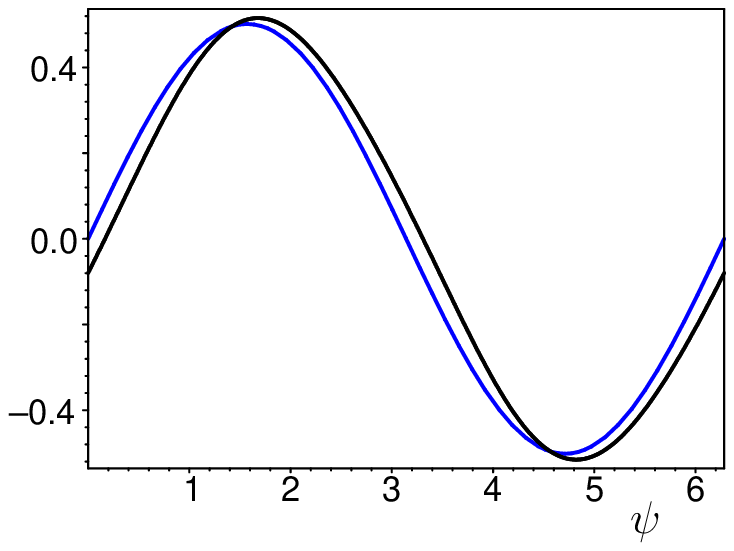}
\\[15pt]
(c)\hspace{-4mm}\includegraphics[width=0.465\textwidth]%
 {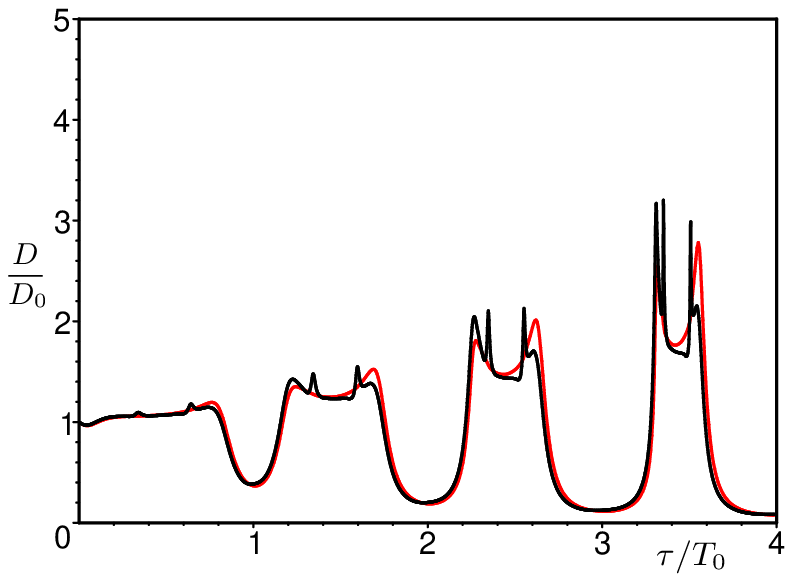}
\\[15pt]
(d)\hspace{-4mm}\includegraphics[width=0.465\textwidth]%
 {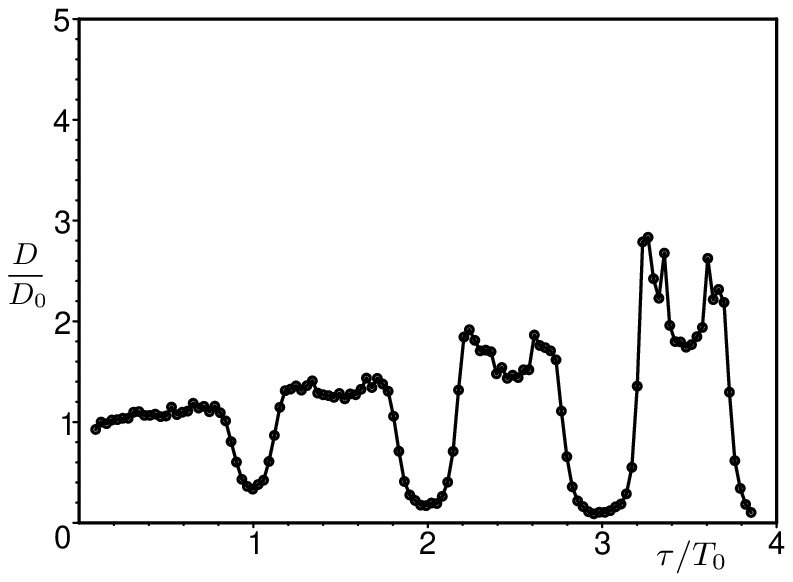}
\end{tabular}
 }

  \caption{Van der Pol oscillator~(\ref{eq-07}) with $\mu=0.7$
subject to the recursive delay feedback control. (a):~The phase
portrait of the system without noise and control. (b):~The
function of susceptibility to feedback $h_{22}(\psi)$ (black
curve) and its harmonic part (blue curve). (c):~The analytical
dependence of the phase diffusion constant on the delay time
$\tau$ (black curve) and the harmonic-approximation version of
this dependence (red curve) for $R=0.5$ and $a=0.1$ ($D_0$ and
$T_0\approx2\pi/0.97$ are the diffusion constant and the mean
oscillation period of the control-free system, respectively).
(d):~Results of a direct numerical simulation with $R=0.5$,
$a=0.1$, noise strength $\varepsilon=0.02$.
 }
  \label{fig1}
\end{figure}
\begin{figure}[!t]
\center{\sf
\begin{tabular}{c}
(a)\hspace{-4mm}\includegraphics[width=0.190\textwidth]%
 {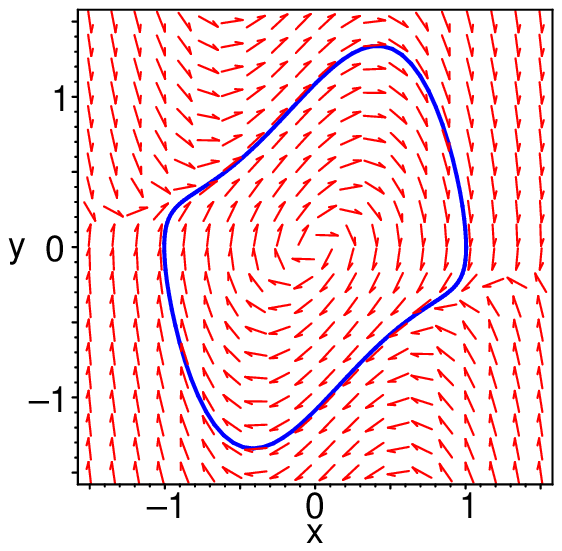}
 \quad
(b)\hspace{-4mm}\includegraphics[width=0.248\textwidth]%
 {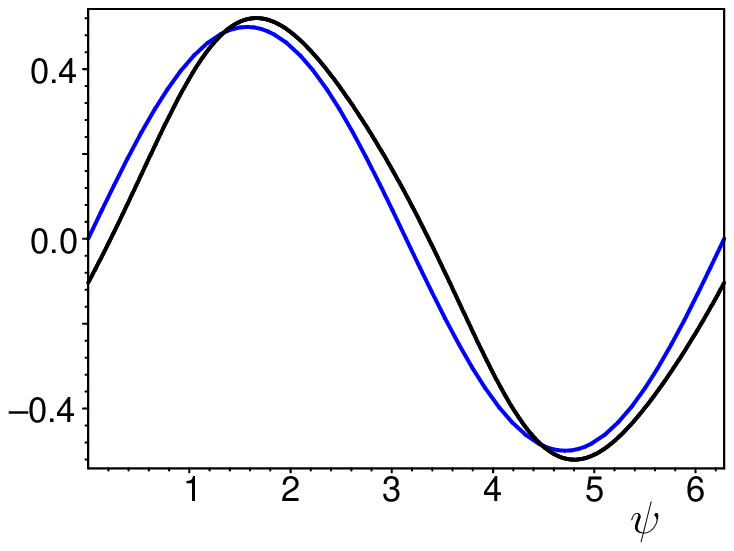}
\\[15pt]
(c)\hspace{-4mm}\includegraphics[width=0.465\textwidth]%
 {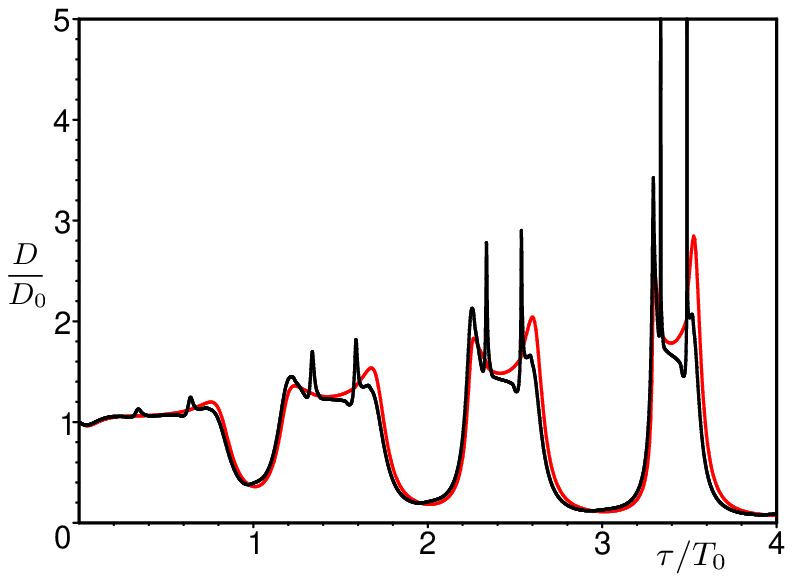}
\\[15pt]
(d)\hspace{-4mm}\includegraphics[width=0.465\textwidth]%
 {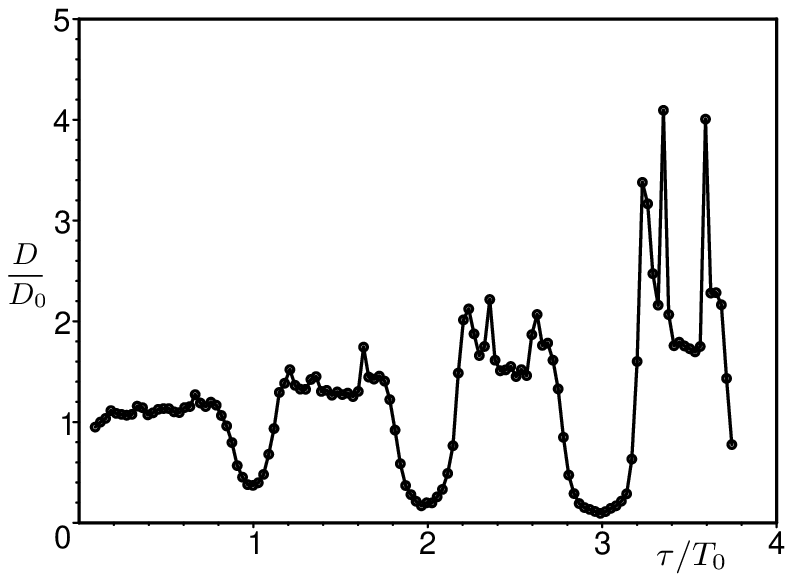}
\end{tabular}
}

  \caption{Van der Pol oscillator~(\ref{eq-07}) with $\mu=1.0$
subject to the recursive delay feedback control. (a):~The phase
portrait of the control-free noiseless system.
(b):~$h_{22}(\psi)$. (c):~Analytical dependence $D(\tau)$
($T_0\approx2\pi/0.94$). (d):~Numerical simulation. For detail see
caption to Fig.~\ref{fig1}.
 }
  \label{fig2}
\end{figure}
\begin{figure}[!t]
\center{\sf
\begin{tabular}{c}
(a)\hspace{-4mm}\includegraphics[width=0.190\textwidth]%
 {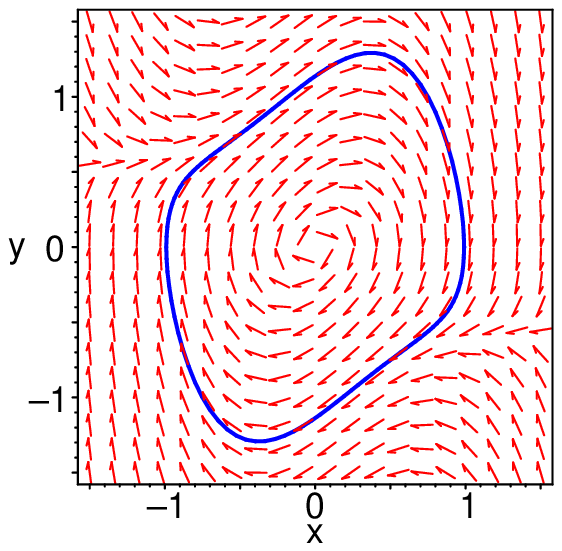}
 \quad
(b)\hspace{-4mm}\includegraphics[width=0.248\textwidth]%
 {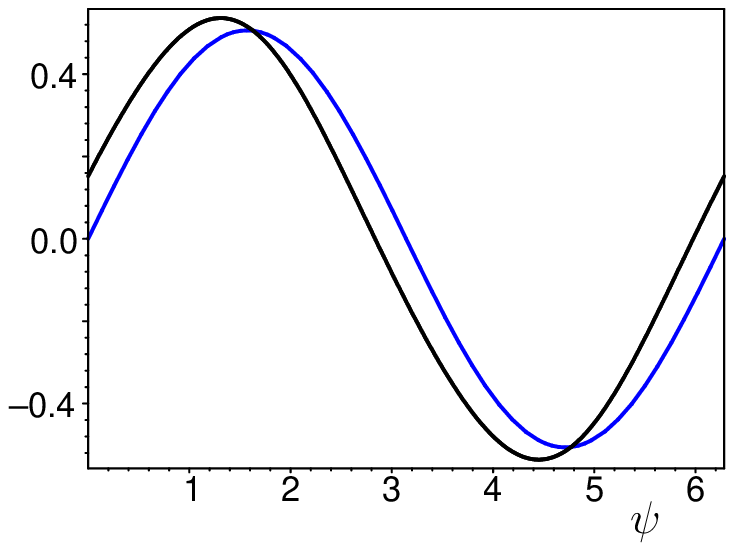}
\\[5pt]
(c)\hspace{-4mm}\includegraphics[width=0.465\textwidth]%
 {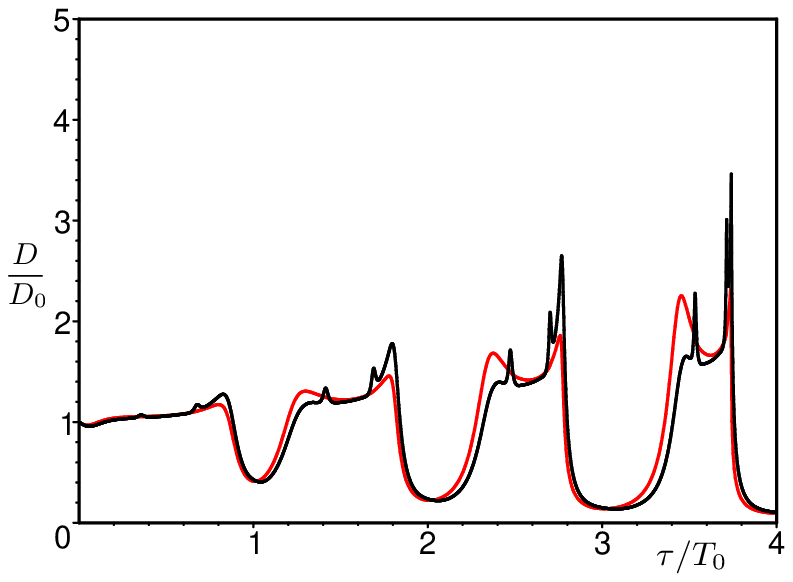}
\end{tabular}
}

  \caption{Van der Pol--Duffing oscillator~(\ref{eq-14})
with $\mu=0.7$ and $b=0.5$ subject to the recursive delay feedback
control. (a):~The phase portrait of the control-free noiseless
system. (b):~$h_{22}(\psi)$. (c):~Analytical dependence $D(\tau)$
($T_0\approx2\pi/1.14$). For detail see caption to
Fig.~\ref{fig1}.
 }
  \label{fig3}
\end{figure}

\section{Recursive delay feedback control for harmonic and anharmonic oscillators}\label{sec3}
\subsection{Harmonic oscillator}
Previous analytical considerations were focused on the case of a
nearly harmonic
oscillators~\cite{Goldobin-Rosenblum-Pikovsky-2003, Goldobin-2008,
Pawlik-Pikovsky-2006, Tukhlina_etal-2008}, such as the van der Pol
one;
\begin{equation}
\dot{x}=y,\quad
\dot{y}=\mu(1-4x^2)y-x+az(t)+\varepsilon\,\xi(t)
\label{eq-07}
\end{equation}
with $\mu\ll1$ ($\mu$ characterizes closeness to the Hopf
bifurcation point). For $\mu\ll1$, the phase
$\varphi=-\arctan(y/x)$ and the limit cycle is: $x_0=\cos\varphi$,
$y_0=-\sin\varphi$; therefore, the equations for
$H_{ij}(\psi,\varphi)$ and $f(\varphi)$ yield
\[
 {\bf\hat h}(\psi)=\left[\begin{array}{cc}
  \frac{1}{2}\sin\psi & \frac{1}{2}\cos\psi\\[5pt]
 -\frac{1}{2}\cos\psi & \frac{1}{2}\sin\psi
 \end{array}\right]\,,\quad
 f(\varphi)=-\sin\varphi\,.
\]
The recursive delay feedback $az(t)$ is typically implemented in
the form
\begin{equation}
z(t)=
 2\sum_{n=0}^{+\infty}R^n\big(y(t-(n+1)\tau)-y(t-n\tau)\big)\,,
\label{eq-08}
\end{equation}
which means
\[
{\bf\hat S}
 =\left[\begin{array}{cc}
 0 & 0 \\
 0 & 1
\end{array}\right],
\]
as we have only the $y$-variable acting on $\dot{y}$ (compare
Eq.~(\ref{eq-07}) and  Eqs.~(\ref{eq-01}), (\ref{eq-02})), and
with $h_{22}(\psi)=(1/2)\sin\psi$ one obtains for a {\it harmonic
oscillator} (cf~\cite{Pawlik-Pikovsky-2006})
\begin{equation}
\Omega=\Omega_0
 -a\frac{(1-R)\sin{\Omega\tau}}{1+R^2-2R\cos{\Omega\tau}}\,,
\label{eq-09}
\end{equation}
\begin{equation}
D=\frac{D_0}{\left(
\displaystyle
 1+a\tau(1-R)\frac{(1+R^2)\cos{\Omega\tau}-2R}{(1+R^2-2R\cos{\Omega\tau})^2}
 \right)^2}\,,
\label{eq-10}
\end{equation}
where $D_0=2\varepsilon^2\langle{f^2}\rangle_\varphi$ is the phase
diffusion constant of the control-free system.

\subsection{Anharmonic oscillator}
Generally, oscillators are anharmonic and $h_{22}(\psi)$ has
additional terms, not only $\sin\psi$. Let us consider
$h_{22}(\psi)$ in the form of a Fourier series;
\begin{equation}
 h_{22}(\psi)=\sum_{m=1}^{\infty}(\alpha_m\sin{m\psi}+\beta_m\cos{m\psi})\,.
\label{eq-11}
\end{equation}
Notice, the term $\beta_1\cos\psi$ belongs to the first harmonic
of the series as well as $\alpha_1\sin\psi$, but it is owed to
anharmonicity; for harmonic oscillations $h_{22}(\psi)$ is purely
proportional to $\sin\psi$.

With Eq.~(\ref{eq-11}), Eqs.~(\ref{eq-04}) and (\ref{eq-05}) take
the form
\begin{eqnarray}
&&\hspace{-12mm}
 \Omega=\Omega_0\nonumber\\
&&\hspace{-10mm}
 -2a\sum_{m=1}^{\infty}\frac{\alpha_m(1-R)\sin{m\Omega\tau}+\beta_m(1+R)(1-\cos{m\Omega\tau})}
 {1+R^2-2R\cos{m\Omega\tau}},
\label{eq-12}
\end{eqnarray}
\begin{eqnarray}
&&\hspace{-12mm}
 D=D_0\Bigg[1+2a\,m\,\tau\,(1-R)\nonumber\\
&&\hspace{-11mm}
 \times\sum_{m=1}^{\infty}
 \frac{\alpha_m[(1+R^2)\cos{m\Omega\tau}-2R]
 +\beta_m(1-R^2)\sin{m\Omega\tau}}
 {(1+R^2-2R\cos{m\Omega\tau})^2}\Bigg]^{-2}\!\!.
\nonumber\\
\label{eq-13}
\end{eqnarray}

\begin{figure}[!t]
\center{\sf
\begin{tabular}{c}
(a)\hspace{-4mm}\includegraphics[width=0.190\textwidth]%
 {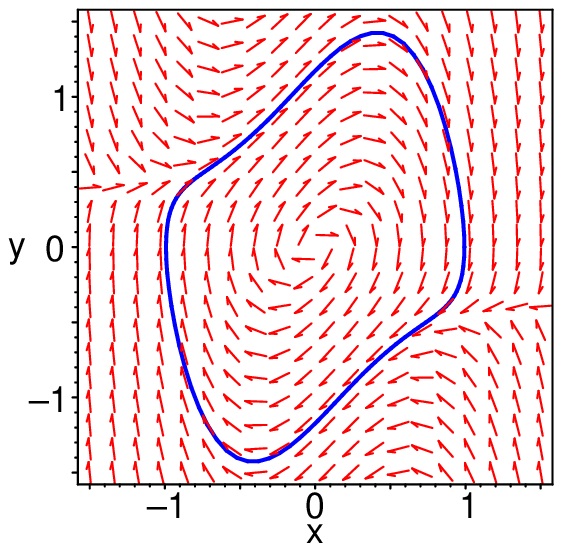}
 \quad
(b)\hspace{-4mm}\includegraphics[width=0.248\textwidth]%
 {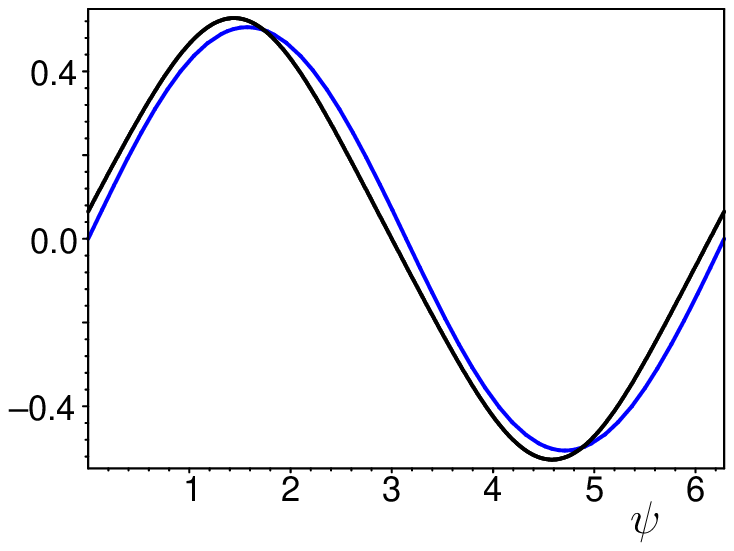}
\\[5pt]
(c)\hspace{-4mm}\includegraphics[width=0.465\textwidth]%
 {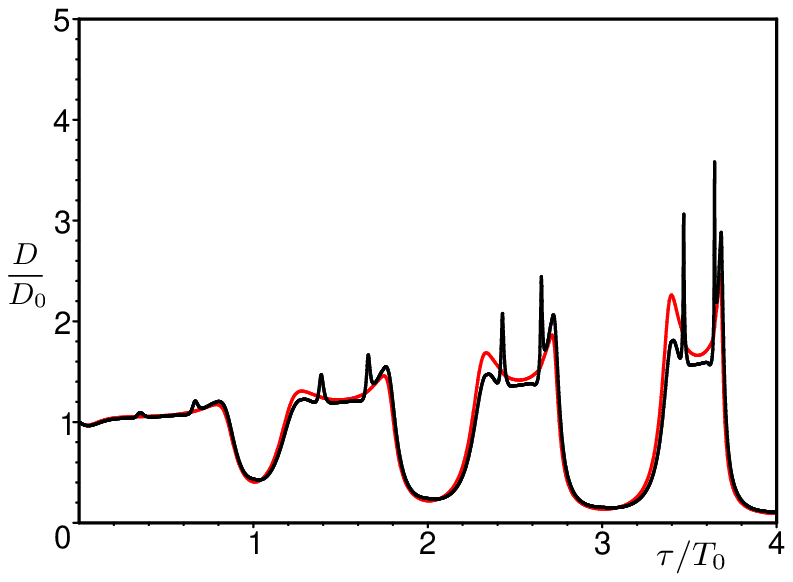}
\end{tabular}
}

  \caption{Van der Pol--Duffing oscillator~(\ref{eq-14})
with $\mu=1.0$ and $b=0.5$ subject to the recursive delay feedback
control. (a):~The phase portrait of the control-free noiseless
system. (b):~$h_{22}(\psi)$. (c):~Analytical dependence $D(\tau)$
($T_0\approx2\pi/1.12$). For detail see caption to
Fig.~\ref{fig1}.
 }
  \label{fig4}
\end{figure}

\begin{table*}[!t]
\caption{Fourier coefficients of $h_{22}(\varphi)$, Eq.~(\ref{eq-11}), for
a van der Pol--Duffing oscillator~(\ref{eq-14}) ($b=0$ corresponds
to a van der Pol oscillator).}
\begin{center}
\begin{tabular}{|c||c|c||c|c||c|c||c|c||}
\hline
&\multicolumn{2}{c||}{$b=0$,   \quad$\mu=0.7$}
&\multicolumn{2}{c||}{$b=0$,   \quad$\mu=1.0$}
&\multicolumn{2}{c||}{$b=0.5$, \quad$\mu=0.7$}
&\multicolumn{2}{c||}{$b=0.5$, \quad$\mu=1.0$}
 \\
\cline{2-9}
 & $\alpha_m$ & $\beta_m$ & $\alpha_m$ & $\beta_m$
 & $\alpha_m$ & $\beta_m$ & $\alpha_m$ & $\beta_m$\\
\hline
 $m=1$ & $+0.501700$ & $-0.076145$ & $+0.499449$ & $-0.093296$
       & $+0.506334$ & $+0.153265$ & $+0.505712$ & $+0.070064$\\
\hline
 $m=2$ & $0$ & $0$ & $0$ & $0$ & $0$ & $0$ & $0$ & $0$\\
\hline
 $m=3$ & $-0.010249$ & $-0.003886$ & $-0.018180$ & $-0.010697$
       & $-0.009246$ & $-0.001205$ & $-0.015938$ & $-0.005048$\\
\hline
 $m=4$ & $0$ & $0$ & $0$ & $0$ & $0$ & $0$ & $0$ & $0$\\
\hline
 $m=5$ & $+0.000044$ & $+0.000084$ & $+0.000035$ & $+0.000422$
       & $+0.000050$ & $-0.000003$ & $+0.000149$ & $+0.000107$\\
\hline
\end{tabular}
\end{center}
\label{tab1}
\end{table*}

\subsection{Anharmonic resonances}
Denominators in the sums in Eqs.~(\ref{eq-12}) and (\ref{eq-13})
have extreme values for $\cos{m\Omega\tau}=\pm 1$. For a single
delay, $R=0$, they equals $1$, while even for moderate values of
$R$ the minima of the denominator can be close to zero.
Remarkably, at points of these resonances ($\cos{m\Omega\tau}=\pm
1$) contribution of the $m$th harmonic can be large even when
$\alpha_m$ and $\beta_m$ are relatively small. For instance, next
to $\tau$ such that $|\cos{3\Omega\tau}|=1$, but
$|\cos{\Omega\tau}|\ne 1$, one can expect resonantly strong or
weak phase diffusion, due to anharmonicity.

For the demonstration and invigoration of the analytical findings,
we have considered a van der Pol--Duffing oscillator:
\begin{equation}
\dot{x}=y,\quad
\dot{y}=\mu(1-4x^2)y-x-bx^3+az(t)+\varepsilon\,\xi(t)\,,
\label{eq-14}
\end{equation}
where $b$ is the Duffing parameter, $b=0$ corresponds to a van der
Pol oscillator (\ref{eq-07}); feedback term $az(t)$ is given by
Eq.~(\ref{eq-08}).

For this oscillator, $h_{22}(\psi)$ and its Fourier transform
(\ref{eq-11}) were calculated. Figs.~\ref{fig1}--\ref{fig4}
present samples of limit cycles (a) and functions $h_{22}(\psi)$
(b); the corresponding Fourier coefficients $\alpha_m$ and
$\beta_m$ are provided in Tab.~\ref{tab1}. (Interested readers can
find a Maple program with comprehensive calculations in the
supplementary material.) One can see in the table, that typical
values of higher harmonics are small: $\alpha_3$ and $\beta_3$ are
only a few percent of $\alpha_1\approx1/2$. Nevertheless, the
dependence $D(\tau)$ exhibits large peaks related to the $m=3$
resonances (in Figs.~\ref{fig1}--\ref{fig4} these dependencies are
plotted for $R=0.5$). The results of calculations without higher
harmonics are plotted with red curves and can be compared to the
one of the ``all-harmonics'' calculation; resonant discrepancies
can be clearly seen. In particular, the $m=3$ resonances are even
stronger than the harmonic ones related to $m=1$.  The results of
a direct numerical simulation show these peaks as well
(Figs.~\ref{fig1}d and
\ref{fig2}d) and are in a good agreement with the analytical
dependencies.

\section{Conclusion}\label{conclusion}
We have considered limit-cycle oscillators subject to weak white
Gaussian noise and recursive delay feedback control. In this way,
we have derived the phase reduction equation and calculated
analytically the mean frequency, the phase diffusion constant, and
the Lyapunov exponent (Eqs.~(\ref{eq-04})--(\ref{eq-06})) for the
case of a general linear feedback (``single'' and recursive delay
feedback, linear frequency filter, etc.) and a general limit-cycle
oscillator.

It has been found that even a small anharmonicity leads to
additional resonances and resonantly strong or weak coherence
(measured by the phase diffusion constant). For instance, for van
der Pol and van der Pol--Duffing oscillators, the anharmonic
resonances due to the harmonic
$h_{22}^{(3)}(\psi)=\alpha_3\sin{3\psi}+\beta_3\cos{3\psi}$ are
even stronger than the harmonic resonances due to
$h_{22}^{(\mathrm{harm})}(\psi)=(1/2)\sin\psi$
(Fig.~\ref{fig1}--\ref{fig4}). Hence, for application of the
recursive delay feedback control, anharmonic resonances are not to
be neglected even for nearly harmonic oscillators.

\appendix
\section{Derivation of Eqs.~(\ref{eq-04}), (\ref{eq-05}), and (\ref{eq-06})}
The derivation presented here is performed in the same way as in
Refs.~\cite{Goldobin-Rosenblum-Pikovsky-2003} for the diffusion
constant and in Ref.~\cite{Goldobin-2008} for the Lyapunov
exponent. Eq.~(\ref{eq-03}) can be rewritten in terms of
$\dot\varphi=\Omega+v(t)$, where $\Omega$ is the mean frequency
and $\langle{v(t)}\rangle=0$;
\begin{eqnarray}
&&\hspace{-3mm}
 \Omega+v(t)=\Omega_0+a\int_0^{+\infty}dt_1
 \sum_{i,j}G_{ij}(t_1)
\nonumber\\
 &&
 {}\times H_{ij}\Big(\Omega(t-t_1)
 +\int^{t-t_1}v(t_2)\,dt_2,\:
 \Omega t+\int^t v(t_2)\,dt_2\Big)
\nonumber\\
 &&\qquad
 {}+\varepsilon f\Big(\Omega t+\int^t v(t_2)\,dt_2\Big)
 \circ\xi(t)\,.\quad
 \label{app-eq-1}
\end{eqnarray}
The mean frequency and robustness quantifiers are related to
long-term behavior of $v(t)$ and, therefore, we can perform
averaging with respect to ``fast'' variable $\Omega t$ (rigorous
multiscale analysis yields the same final expressions but in a
very lengthy way and is not presented here). Eq.~(\ref{app-eq-1})
turns into
\begin{eqnarray}
&&\hspace{-3mm}
 \Omega+v(t)=\Omega_0+a\int_0^{+\infty}dt_1
 \sum_{i,j}G_{ij}(t_1)
\nonumber\\
 &&
 {}\times h_{ij}\Big(-\Omega t_1
 -\int_{t-t_1}^t v(t_2)\,dt_2\Big)
\nonumber\\
 &&\qquad
 {}+\varepsilon f\Big(\Omega t+\int^t v(t_2)\,dt_2\Big)
 \circ\xi(t)\,,\quad
 \label{app-eq-2}
\end{eqnarray}
where $h_{ij}(\psi):=\langle
H_{ij}(\varphi+\psi,\varphi)\rangle_\varphi$,
 $\langle...\rangle_\varphi\equiv(2\pi)^{-1}\int_0^{2\pi}...\,d\varphi$.
Linear in $v$ approximation reads
\begin{eqnarray}
&&\hspace{-3mm}
 \Omega+v(t)=\Omega_0+a\int_0^{+\infty}dt_1\sum_{i,j}G_{ij}(t_1)
\nonumber\\
 &&
 {}\times\Big[h_{ij}(-\Omega t_1)
 -h_{ij}'(-\Omega t_1)\int_{t-t_1}^t v(t_2)\,dt_2\Big]
\nonumber\\
 &&\qquad
 {}+\varepsilon\Big[f(\Omega t)+f'(\Omega t)\int^t v(t_2)\,dt_2\Big]
 \circ\xi(t)\,.\quad
 \label{app-eq-3}
\end{eqnarray}

Averaging Eq.~(\ref{app-eq-3}) yields
\begin{equation}
\Omega=\Omega_0+a\int_0^{+\infty}
  \sum_{i,j}G_{ij}(t)\,h_{ij}(-\Omega t)\,dt+O(\varepsilon^2)\,.
\label{app-eq-4}
\end{equation}
Further, multiplying Eq.~(\ref{app-eq-3}) by $\xi(t+s)$ and
averaging, we can evaluate $\langle\xi(t+s)\,v(t)\rangle$ from
known $\langle\xi(t+s)\,\xi(t)\rangle$. Multiplying the same
equation by $v(t+s)$ and integrating, we evaluate
$\langle{v(t+s)\,v(t)}\rangle$ from
$\langle{v(t+s)\,\xi(t)}\rangle$. The integral of the
autocorrelation function,
$\int_{-\infty}^{+\infty}\langle{v(t+s)\,v(t)}\rangle\,ds$, yields
the phase diffusion constant $D$;
\begin{equation}
D=\frac{2\varepsilon^2\langle{f^2}\rangle_\varphi\,\big(1+O(\varepsilon^2)\big)}
 {\Big(1+a\int_0^{+\infty}t\sum\limits_{i,j}
 G_{ij}(t)\,h'_{ij}(-\Omega t)\,dt\Big)^2}\,.
\label{app-eq-5}
\end{equation}

Similarly to Ref.~\cite{Goldobin-2008} we find that the leading
Lyapunov exponent
\begin{equation}
\lambda=-\frac{\varepsilon^2\langle(f')^2\rangle_\varphi\,\big(1+O(\varepsilon^2)\big)}
 {\Big(1+a\int_0^{+\infty}t\sum\limits_{i,j}
 G_{ij}(t)\,h'_{ij}(-\Omega t)\,dt\Big)^2}\,.
\label{app-eq-6}
\end{equation}

Eqs.~(\ref{app-eq-4})--(\ref{app-eq-6}) with $\varepsilon\ll1$
turn into Eqs.~(\ref{eq-04})--(\ref{eq-06}). Remarkably, the ratio
$\lambda/D$ is constant only in the leading order with respect to
$\varepsilon$; the proportionality between $\lambda$ and $D$ does
not hold for higher order corrections.


\end{document}